\documentstyle[twoside,psfig,fleqn]{article}

\def\P{\rm{\bf Prob}}
\def\endproof{\hfill\vrule height .6em width .6em depth 0pt\goodbreak\vskip.25in }

\title{The Anderson Model as a Matrix Model}

\author{
J. Magnen, G. Poirot, and V. Rivasseau\\
Centre de Physique Th\'eorique, Ecole Polytechnique
\\
F91128 Palaiseau Cedex}

\begin{document}

\maketitle
\vskip 1cm
\begin{abstract}
In this paper we describe a strategy to study the Anderson model
of an electron in a random potential at weak coupling 
by a renormalization group analysis. There is an interesting 
technical analogy between this problem and the theory of
random matrices. In $d=2$ the random matrices which appear are 
approximately of the free type well known to physicists and mathematicians,
and their asymptotic eigenvalue distribution is therefore simply Wigner's law.
However in $d=3$ the natural random matrices that appear 
have non-trivial constraints of a geometrical origin.
It would be interesting to develop a general
theory of these constrained random matrices, 
which presumably play an interesting role for
many non-integrable problems related to diffusion.
We present a first step in this direction, namely
a rigorous bound on the tail of the
eigenvalue distribution of such objects
based on large deviation and graphical estimates. This bound allows to prove
regularity and decay properties of the averaged Green's functions 
and the density of states for a three dimensional 
model with a thin conducting band 
and an energy close to the border of the band, for sufficiently small
coupling constant. 

\vskip 1cm

This paper is dedicated to Claude Itzykson.

\end{abstract}



\section{Introduction and Warning!}
The reader might be slightly surprised (perhaps even baffled?) by this
contribution. Indeed it has nothing to do directly
with the main topic of these Proceedings, namely 
strings, membranes and duality. It is however a tradition (and a good
one!) to include in this cycle of conferences some applications
of quantum field theory to condensed matter physics. But in the case of our
contribution, not only its subject, the
mathematical analysis of disordered
systems, is quite marginal from the point of view of
the main theme of the conference, but also its form.  It is not a 
``pedagogical introductory review''
on the subject (for such a review we refer e.g. to [A][S]). We chosed indeed 
to give here the main technical elements of the proof of a 
non-perturbative regularity theorem 
on the Anderson model of an electron scattered by a random
potential. We hope that glancing at this paper might nevertheless be amusing
or even useful to the reader for two reasons.
First the arid details of the mathematical bounds below
will remind us how much effort is necessary
to transform into proof even what 
certainly looks as the most trivial 
of all theoretical issues. Second we hope
that the reader might be interested by the tool that was recently
introduced in this area, namely a new category
of random matrices which have constraints of a geometrical origin. 
The free random matrices
with independent entries considered by Wigner, Dyson and others (and
to the theory of which Claude Itzykson contributed so successfully)
held the key to many unexpected theoretical developments which range
from nuclear physics and confinement to quantum gravity
and non-perturbative strings. There is now a general mathematical frame
for such matrices, namely the free non commutative probability theory
of Voiculescu [V]. In this frame, Wigner's law appears very general:
it is simply the non commutative
analog of the Gaussian law of large numbers for
independent ordinary (commutative) random processes. But the theory of 
{\it constrained} non commutative processes remains to be developed.
There are many three dimensional problems (the BCS theory, the
Anderson model, scattering of particles in 3 or 4 dimensional Minkovski space)
that can be discretized as problems
of random matrices with geometrical constraints of the type considered 
below in section 4. These matrices
seem definitely to have quite different scaling properties from the usual ones
with independent entries.
Perhaps they might also become useful tools in many other domains, 
when integrability and two
dimensional conservation laws are no longer available. Therefore we
think that the scaling properties of such random matrices of non-free type 
deserve to be further investigated, even numerically. 

\section{The Anderson Model} 

The Anderson model of a single electron in a random potential 
was introduced to study the concept of localization. 

Rigorous results in dimensions not restricted to 1 up to now concern
the localization regime. The first result of this kind was
the multiscale analysis of Fr\"ohlich and Spencer, who
proved absence of diffusion for large disorder or energy 
out of the conduction band [FS].
It was later proved that in this regime localization holds,
namely the expectation value of the modulus
of the Green's function decays exponentially
and the spectrum of the Hamiltonian is almost surely pure
point [DLS], [FMSS]. The proofs have been simplified more recently 
by an argument due to Aizenman and Molchanov [AM], which is based
on considering the
expectation value of the modulus of the Green's function to a power
strictly smaller than one.

However almost nothing rigorous
is known about the ``extended states'' 
regime where the energy lies in the middle of the spectrum
of the free Hamiltonian and the disorder is weak. This is the regime
where diffusive conduction should take place, at least in 3 dimensions.
There are only results at $d=\infty$ by Klein [K]. 

We think that the program of renormalization group around singular
manifolds should be useful to study this regime.
This program was originally developed for the study of many Fermions 
and in particular for a rigorous understanding of the BCS theory
of superconductivity [FT], [FMRT1]. In that context, the singular manifold
was the Fermi surface. In the case of the Anderson model,
the singular manifold is simply defined by the equation $p^{2}=E$,
namely it is a sphere for a free rotation invariant Hamiltonian;
in the case of a lattice model breaking rotation invariance, one would have
a slightly more complicated manifold such as e.g.
$\sum_{i=1}^{d} (2-2\cos p_{i}) = E $.

This renormalization group analysis consists in slicing the free Green's
function around the singular locus in momentum space, and performing a phase
space analysis (by means of cluster expansions) which allows to compute the
long range behavior governed by the effective theory near the singularity.

In order to control this renormalization group analysis it is useful
to understand the behavior of the theory in a single momentum slice,
hence slightly away from the singularity,
and to prove that perturbation theory in this case is meaningful
(i.e. asymptotic), using the fact that the coupling is weak to control
the cluster expansion. 

In a previous paper [P] this single slice analysis was performed in two
dimensions. In that situation the main problem consists in
perturbing the identity matrix by a random matrix with independent
coefficients, and this problem can be controlled
using e.g. the asymptotics to the Wigner's law for the density
of eigenvalues of this kind of matrices. 

In three dimensions the analogous problem is more difficult
since the matrix no longer has independent random entries.
It belongs to this new class of random matrices alluded above, for which  
large size asymptotics are not known. In this paper we initiate
the study of such matrices by large deviation techniques which ultimately
rely on Feynman graphs estimates. We prove a theorem which states that
the probability that such random matrices develop a large eigenvalue is
sufficiently small, so that a single slice analysis again can be
controlled.

This single slice analysis can be considered either as a step in the program
to analyze by the RG method the situation where the energy lies in the middle
of the conduction band, or as a full result in its own right, but
for an energy out of the conduction band
and close to its border. In this last case our probabilistic bounds
imply via a cluster expansion 
that the averaged Green's function has fast decay
(for a $C^{\infty}_{0}$ band cutoff, this decay is polynomial of 
arbitrarily high order), and that it is asymptotic to its
perturbative expansion.


\section{Notations and Results}

\subsection {Definitions}

We consider an electron in a weak random potential in three dimensions. 
The mass of the electron is set to $m=1/2$ so that in these units
the ordinary kinetic energy in Fourier space is simply $p^{2}$.
To model a conduction band, we consider a free Hamiltonian $H_{0}$
which is simply a multiplication operator in Fourier space:
$$
\widehat H_{0} (p) = p^{2}/\eta(p) \eqno(1)
$$ 
where $\eta$ is some rotation invariant
$C_{0}^{\infty}$ function which is 1 on a certain
interval $[a,b]$ and zero on an interval $[a',b']$, with $a'>0$. Therefore
in this model $H_{0}$, the energy, is both infinite for
large momenta $|p|>b'$ and for small momenta ($|p|<a'$), so there is
both an ultraviolet cutoff, as should be the case
for any non-relativistic problem, and an infrared cutoff on small
momenta.
Lattice cutoffs could also be treated
by our methods, replacing for instance the function $p^{2}$ 
by $\sum_{i}(2-2\cos p_{i})$ (hence they are not rotation invariant).

The interacting Hamiltonian is obtained by adding to $H_{0}$
a random potential $V$:
$$ H = H_{0} + \lambda V  \eqno(2)
$$ 
We take $V$ as a Gaussian ultra-local field in $x$-space, {\it i.e.} each
$V(x)$ is an independent Gaussian variable. More precisely, it means that $V$
is distributed according to a normalized Gaussian measure $d\mu_{\kappa}$
of covariance $\kappa$. We also include a volume cutoff: this means 
that we choose a large box $\Lambda$ of volume $|\Lambda|$
and side size $L=|\Lambda|^{1/3}$, and restrict the operator
$V$ to that box. We can also use e.g. periodic boundary conditions
for the operator $H_{0}$ on the boundary $\partial\Lambda$ (although
this last condition is not essential). In that case the Fourier-transform
of a $\Lambda$-periodic function $f(x)$ is
$$
   f(p)=\int_{\Lambda} e^{ip.x} f(x) \, dx 
\eqno(3)$$
and verifies
$$   f(x)=\frac{1}{|\Lambda|} \sum e^{-ip.x} f(p) \ \mbox{with }
   p \in (\frac{2\pi\mbox{\sf Z\hspace{-5pt}Z}}{L}) ^{3} \eqno(4)
$$

\setcounter{equation}{4}

We are interested in computing the averaged Green's function at energy $E$
(with a Feynman prescription)
as a kernel in $x$-space, or the density of states, which is the 
imaginary part of this kernel at coinciding points. 
Therefore we want to analyze
\begin{eqnarray}
 < G_{\pm}> &=&  \lim_{\Lambda
    \rightarrow \infty} \lim_{\varepsilon \rightarrow \pm 0} 
   < G_{\Lambda, \varepsilon}> 
\end{eqnarray}
\begin{equation}
  <G_{\Lambda, \varepsilon}>=\int \! 
      \frac{1} {H_{0}+\lambda V -(E+i\varepsilon)} \, d\mu_{\kappa}(V)
\end{equation}
In the following we will note $<.>$ for integration with respect to $V$,
and $\P$ for the probability of an event, measured with respect to
$d\mu_{\kappa}$.

It is important to remark that we can put some kind of
ultraviolet cutoff on the random potential $V$ as well,
because we are treating a non-relativistic problem. Usually this is done
by considering the potential on a lattice, but we can also choose 
to stay in the continuum and to
take the covariance $\kappa$ of $V$ not to be a $\delta$
distribution but a smooth positive type kernel with compact support or
fast decay at the unit scale (the scale of our ultra-violet cutoff).

\subsection{Expected Results}

In the standard Anderson model at weak coupling in dimension 3 the first 
significant open mathematical problem is to prove that for a conduction band
of finite width and an energy $E$ inside the band (that is $E \in [a,b])$
the thermodynamic limit of Green's functions defined above exist and 
has fast decay (exponential decay in the case of the lattice cutoff)
for real small enough coupling $\lambda$. One want also to prove
that the decay rate behaves, as a function of the coupling, as:
$$
  \Delta =\pi^{2} \lambda^{2} + O(\lambda^{4}) \eqno(7)
$$

All these conjectures stem from the well-known perturbative analysis
of the model. It is based on expanding (6) as a power series
in $\lambda$ (resolvent expansion), then integrating on $V$.
This perturbative analysis is of course plagued by the usual divergence of
perturbation theory.

To obtain the leading behavior $\Delta \simeq \pi^{2} \lambda^{2}$,
it is sufficient to compute a self-consistent renormalization of the
divergent tadpole graph, which leads to:

$$ \Delta \simeq \lambda^{2} \Im \int d^{d}p {\eta(p) 
\over p^{2} -( E +i \Delta )}
\eqno(8)
$$
so that 
$$  
 1 \simeq \lambda^{2} \int d^{d}p { \eta(p) 
\over( p^{2} - E)^{2} + \Delta^{2} } \ .
\eqno(9)
$$
This equation gives the behavior (7) since $E$
belongs to $[a,b]$, where the cutoff $\eta$
is 1. This self-consistent
computation is the  so-called ``rainbow graphs'' approximation.

Thus the first main effect of $V$ is to add a mass term in $O(\lambda^{2})$
which screens the singularity of the free propagator $\displaystyle
\frac{1}{p^{2}-E}$. Physically this means that the electron wave function
loses memory of its phase by the incoherent reflections on $V$.

To prove this fast decay of $<G_{\pm}>$
or equivalently to analyze the behavior at small $\lambda$ of the density
of states is our first natural goal in the small coupling analysis of 
the Anderson model. Remark however that
this result would still fail to distinguish
between $localization$ or the existence
of $extended$ $states$, which can be 
distinguished only by the decay properties
of $<G_{+}G_{-}>$. Let us recall that in the localization regime 
(which in $d=3$ means for $\lambda$ large) the function 
$<G_{+}(x,y)G_{-}(x,y)>$
decays exponentially, whether in the diffusive or extended states regime
(which in $d=3$ should exist for $\lambda$ small),
it is conjectured that the function $<G_{+}(x,y)G_{-}(x,y)>$
decays only as $|x-y|^{-1}$. To distinguish rigorously between these
two regimes at $d=3$, hence to establish the existence of the Mott-Anderson
transition is the long term goal for our program.

\subsection{A Result}

Here we do not even prove yet
this fast decay of $<G_{\pm}>$ when $E$ is inside the band,
but establish a weaker result. 
We consider an energy $E$
not inside the band but near its edge. Also the coupling constant $\lambda $ 
is taken real and small enough so that both 
the width of the band $b'-a'$ and the distance between
the energy and the border of the band (say, $a'-E$ for an energy 
close to the {\it lower} edge of the band)
are large compared to $\lambda^{2}$.
Under these conditions we shall prove that the averaged
Green's function has fast decay  on a spatial
scale of order $r^{-1}$, where $r= \inf \{b'-a', a'-E\}$. 
More precisely we can state the:

\medskip
\noindent{\bf Theorem I}
{\it Let $r= \inf \{b'-a', a'-E\}$.
There exists some small constant $\epsilon$ such that if 
$\lambda^{2} \le \epsilon r$, then $<G_{\pm}>$
exist (the convergence of its kernel
as $\Lambda\to \infty$ and $\varepsilon \to 0$ 
being uniform on all compact sets) and for all integer $m$ there exists
a constant $K_{m}$ such that:
$$ |<G_{\pm}>(x,y)| \le K_{m} {1 \over (1 + r |x-y|)^{m}}\ .
\eqno(10)
$$
Furthermore this averaged Green's function is asymptotic to all orders
its perturbative expansion in $\lambda$. This expansion is obtained
by expanding the resolvent (see (19) below)
and performing the Gaussian integration
on $V$ in $<G_{\pm}>(x,y)$.}

\medskip
This result in $d=2$ is essentially the subject of [P]. The purpose
of these notes is to extend it to $d=3$, and explain the difference.
The proof combines two aspects, a large/small field probabilistic analysis
which is given here in detail, and a cluster expansion which is
a rather standard method ([R][P]), so not repeated here.

\subsection{Next steps...}

The same result can
be proved to hold also 
for a  band width no longer small but finite, of order 1, 
the distance of the energy to the border of the band being still small
(hence $\lambda^{2} \le \epsilon
r = a'-E$); since the proof is technically slightly
more difficult, we do not include it here.

Finally the much more difficult problem of an energy {\it within} the band,
is somewhat analogous to spontaneous mass generation in field theory or in 
the BCS theory. Our result which concerns
the energy outside the band 
is nevertheless a relevant step towards this more difficult result,
since one can expect to analyze the case of the energy inside the band
by cutting slices closer and closer to it, as in a renormalization group
analysis [P]-[FT].
We must however add that the situation is quite different
in dimensions 2 and 3. In two dimensions we 
have a precise scenario in progress to solve this problem
of an energy within the band,
using the specific two dimensional
rules of conservation of momentum [MPR]. In
three dimensions we have not yet such a precise scenario on how to 
attack the problem of an energy within the band.


\subsection{Complex translation and limit $\varepsilon
\rightarrow 0$} 
We can make an {\em a priori} partial renormalization and 
generate a non vanishing imaginary part in the denominator thanks to 
a complex translation of the integration contour of $V$, using the formulas
$$
      \int d\mu_{C}(X) \, {\cal F}(X) \ = \quad e^{( aCa)/2} 
$$
$$
         \int d\mu_{C}(X) \, {\cal F}(X+iCa) \, e^{-iaX} \ .
\eqno(11)   $$
These formulas are 
easily proved for ${\cal F}$ polynomial and then extended with a
density argument to meromorphic functions provided we meet no pole.

Let us translate $V$ by $\displaystyle -i \frac{\delta}{\lambda} 1$, and get:

\setcounter{equation}{11}
\begin{equation}
  G_{\Lambda, \varepsilon} = e^{|\Lambda| \delta^{2}/2\lambda^{2}}\left<
    \frac{e^{i\delta1.V/\lambda}}{H_{0} - (E+i\delta + i
    \varepsilon) + \lambda V} \right> 
\end{equation}
(The exact value
of $\delta$ will be fixed later on).
Now we can take the limit $\varepsilon \rightarrow 0$
which yields
\begin{equation}
  G_{\Lambda} = e^{|\Lambda| \delta^{2}/2\lambda^{2}} \left<
    \frac{e^{i\delta 1.V/\lambda}}{H_{0} - (E+i\delta) +\lambda V}
    \right>\ .
\end{equation}
On the lattice this formula is exact but 
in the continuum it has to be slightly corrected
to take into account
the  ultraviolet cutoff on $V$, but this
is inessential [P].

The cost of this operation is the bad factor $e^{|\Lambda|
\delta^{2}/2\lambda^{2}}$ that we may have to compensate
if some absolute value in the bounds cancels the oscillating 
term $e^{i\delta 1.V/\lambda}$. However let us explain the strategy:
space will be divided into so-called large/small field regions.
In the small field regions, perturbation theory works. The eigenvalues
of $\lambda V$ are too small to cancel $H_{0}-E$. Therefore the imaginary term
$i\delta$ is not really needed and the complex translation
can be reversed, so that the bad factor has not to be compensated.
In the large field region, this is no longer true, but the bad factor
is more than compensated by the probabilistic factor associated to the
measure $d\mu_{\kappa}$ which is very small.


\subsection{The resolvent expansion}

From now on the number $r$ is chosen of the form 
$r=M^{-j}$, where $M$ is some fixed number and $j$ a large positive integer.
These are the usual notations which come from
the renormalization group point of view
[FT],[FMRT][P]. We also set $E=1$ for simplicity.
Let us also define $\eta (p) = [\eta _{j}(p)]^{3}$ 
(the subscript $j$ simply recalls
that the width of the support of the function $\eta$ is $M^{-j}$). 
We introduce the free Green's function
$$
    \Gamma = (H_{0}-(1+i\delta))^{-1} =
\eta_{j} C \eta_{j}\ ; \eqno(14)
$$
where
$$ C= {\eta_{j} \over p^{2} - \eta_{j}^{3} (1+i\delta)} \ . \eqno(15)
$$
Then we can factorize $C$ and rewrite $G$ as  
$$
   G_{\Lambda} = e^{|\Lambda| \delta^{2}/2\lambda^{2}}
      \Bigg< e^{i\delta 1.V/\lambda} \, \eta_{j} 
\frac{1}{1+\lambda C\eta_{j} V \eta_{j}} C \eta_{j} \Bigg> 
$$
$$
\quad \quad \quad \quad \quad \quad \equiv {\cal A}
      \left< e^{i\delta 1.V/\lambda} {\cal R}(V) \right> \eqno(16)
$$
where we used the operator product conventions
$$
   (A B)(x,y)=\int A(x,z) B(z,y) \, dz 
$$
$$ \Leftrightarrow \quad 
   (A B)(p,q)=\frac{1}{|\Lambda|} \sum A(p,k) B(-k,q) \eqno(17)
$$
where
$$
   A(p,q)=\int e^{ip.x-iq.y} A(x,y) \, dx \, dy\ . \eqno(18)
$$

Because of its smooth covariance $V$ is almost surely bounded. Thus for a
given $V$, ${\cal R}(V)$ is analytic in $\lambda$ in a small domain around the
real axis. This implies that $\cal R$ is the unique analytic continuation
of the operator series
\setcounter{equation}{18}
\begin{equation}
  {\cal S} = \sum_{n=0}^{\infty} (-\lambda)^{n} \eta_{j}
  (\underbrace{ CH C \ldots  H C}_{n
  \scriptsize H' s} ) \, \eta_{j}
\end{equation}
\begin{equation}
{\rm with}\ \ \   H = \eta_{j} V \eta_{j} \ .
\end{equation}

\section{Phase space Analysis}
We want to study $G$ by means of a {\em phase space analysis}. 
In order to do so,
we will cut the propagator $C$ according to cells (called sectors)
in the space of angular coordinates
of the momentum $p/|p|$ (see subsection  below).
These sectors are also the main tool in the constructive
renormalization group analysis of [FMRT1-4].
Space will also be cut according to a lattice of cubes
of size $M^{j}=R^{-1}$, which is adapted to the decay of the propagator,
or of the cutoff $\eta_{j}$. In momentum space
$V$, seen as an operator, is $V(p,q) \equiv V(p-q)$.

\vskip 1cm
\centerline{\hbox{\psfig{figure=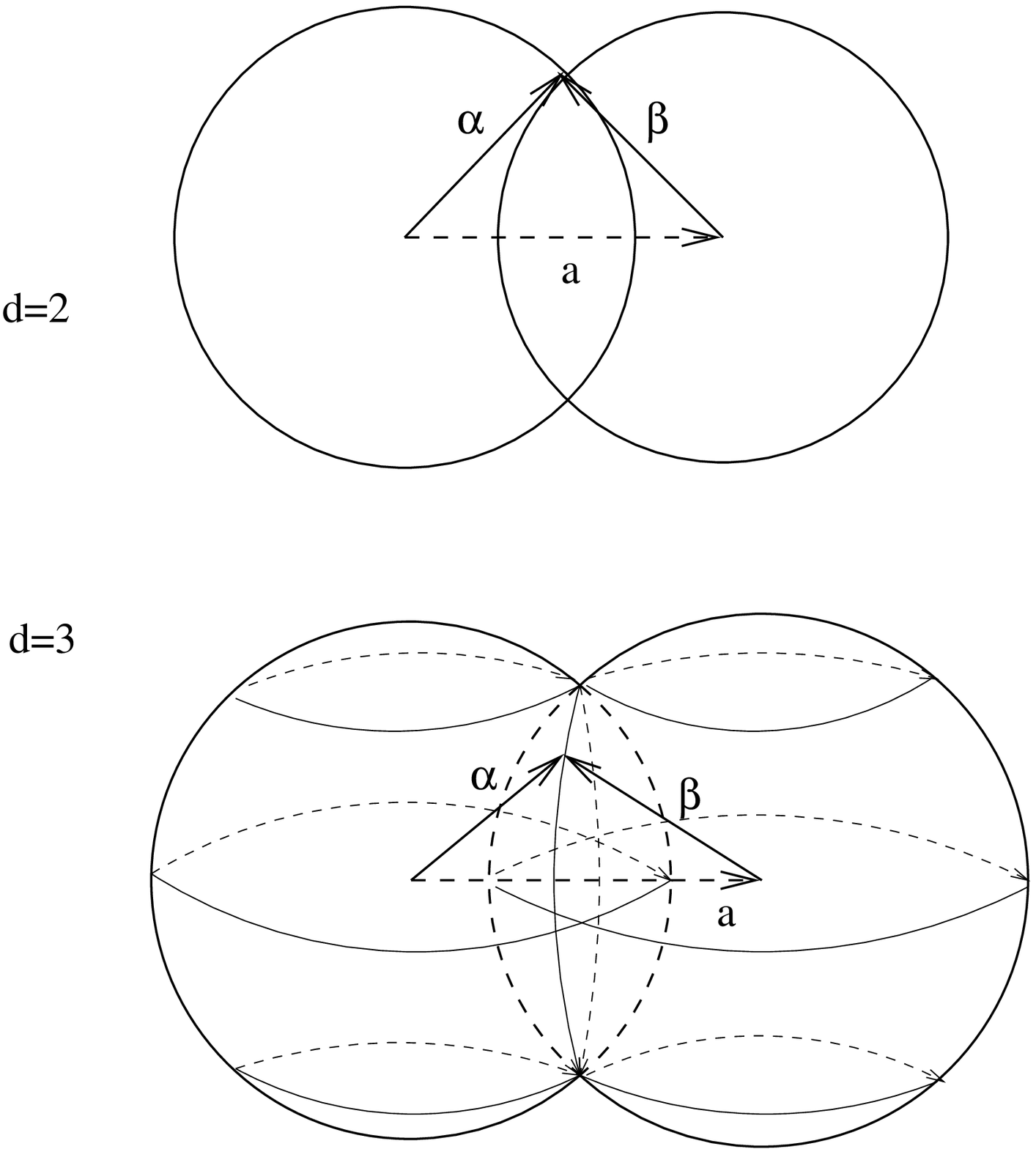,height=7.5cm,width=7cm}}}

\vskip 1cm 
\centerline{\bf Figure 1}
\medskip

The key difference
between the constructive or phase space analysis in 2 and 3 dimensions is that
in two dimensions when $p$ and $q$ have the same fixed norm 
(here close to $E=1$, since $|p| -1 \le 2r = 2M^{-j}$), the sum $p+q$
defines in a unique way the pair $\{p,q\}$, whether in three dimensions
it defines it only up to a rotation by an angle (see Figure 1).
The operator $V$ can be considered as a matrix whose entries are labeled by
this discrete set of sector indices. In this point of view, the
geometrical behavior depicted in Figure 1 translates essentially in
the following property: the matrix 
obtained in $d=2$ is approximately an orthogonal ensemble with Wigner's
semi-circle law as asymptotic behavior of the eigenvalues density [P],
whether in three dimensions the matrix is of a constrained
type, and the asymptotic distribution for the eigenvalues can be
probed only more indirectly, for instance, as will be shown below,
through large order deviations which amount
to the computation of particular Feynman graphs. We go on
to make now this statement more precise.


\subsection{The discretized ``single cube problem'' as a Random Constrained
Matrix}

The usual point of view in localization theory is to treat space as 
fundamental and to expand 
the off diagonal part of $C$, which couples different points. But for
diffusion theory and extended states regimes, it is better to keep
momentum space as fundamental.
Since $C$ is a diagonal operator in momentum space, the important 
part of the problem is to understand the behavior of the random
operator $H$ in momentum space. We show now how to
discretize this operator as a matrix to understand better its 
qualitative behavior.

From now on we restrict ourselves to dimension $d=3$. 
Suppose that we cut the momentum shell support
of  the cutoff $\eta_{j}$ into cells $\sigma$ (isotropic sectors)
which have size $M^{-j}$ in all directions:
$\eta_{j} =\sum_{\sigma} \eta_{j,\sigma}$. There will be roughly
$N= M^{2j}$ such sectors (neglecting factors such as $4\pi$). 
Consider the covariance $\kappa$ of $V$. Since $V$ is almost ultralocal,
this covariance is almost 1 on a ball of large radius. But by momentum
conservation, the operator $H_{\sigma,\sigma'}
=\eta_{j,\sigma}V\eta_{j,\sigma'}$ is zero unless the momentum
of $V$, which should be approximately 
$\sigma -\sigma '$,
lies inside the ball of radius $2 +4r$, so approximately the ball of
radius 2. 
In this ball 
$\kappa (p)=1$ with a very good approximation. Cut
the interior of this three dimensional
ball of radius 2 into about $M^{3j}\simeq N^{3/2}$ cells $c$ of side size
$M^{-j}$: $\kappa = \sum_{c} \kappa_{c}$. 
There will be an associated decomposition of $V$ as a 
sum of random variables $V_{c}$, each independent and 
distributed with covariance $\kappa_{c}$.

When this decomposition of momentum space both for the electron cutoff
$\eta_{j}$ and for $V$ has been made 
with smooth $C_{0}^{\infty}$ functions, there is fast
dual spatial decay of the propagators $\eta_{j,\sigma}$ 
and $\kappa_{c}$ on length
scales $M^{j}$. Therefore the heart of the theory is given by the problem
restricted to a single spatial cube of side size $M^{j}$. To understand this
single cube theory we can neglect now the space dependence 
of $\eta_{j,\sigma}$ and $\kappa_{c}$, keeping only the constraint 
of momentum conservation. 

In this approximation the
different random variables $V_{c}$ (which are no longer fields but real 
random variables) become identically distributed.
The matrix $H$ has now $N^{2}$ entries, namely all pairs $(\sigma, \sigma')$
where $\sigma$ and $\sigma'$ are isotropic sectors. We understand that since
this $N$ by $N$ matrix depends on $N^{3/2}$ independent variables only, 
it cannot be of the free type usually considered in theoretical physics
(orthogonal, hermitian or symplectic ensemble). It
is of a new {\it constrained} type, and the constraints stem from
the geometry of momentum addition on the singular manifold
(here the sphere). These constraints are related to 
the absence, in dimension 3 or more, of conservation rules for diffusion and of
completely integrable models.
Because of these constraints, 
the matrices obtained are not invariant under a group
such as the orthogonal, unitary and symplectic group; the statistics
do no longer factorize and reduce to eigenvalue statistics (namely there
are ``preferred eigenvectors'' as well), orthogonal polynomial methods 
do not apply, etc..., and nothing seems to be 
known up to now about such matrices.

These matrices can nevertheless be  
described and studied numerically on a computer very easily
with the following program (here we simplified again slightly the initial
problem, which involved complex matrices, replacing it by an
analogous real problem with sums of momenta instead of differences):

Take a small number $r=M^{-j}$. Divide the unit sphere into $N\simeq M^{2j}$
sectors of approximately similar area $4\pi r^{2}$. Divide the ball
of radius two into $N^{3/2} \simeq M^{3j/2}$ cubic cells of volume
approximately $(4/3)8\pi r^{3}$. Generate with a random number generator
$N^{3/2}$ independent random variables $V_{c}$, one for each cell
of the ball. Associate to any pair of sectors $(\sigma, \sigma')$
the unique cell $c=c(\sigma, \sigma')$ such that the sum of the two momenta
at the center of $\sigma $ and $\sigma '$ fall into the cell $c$, and
plug the corresponding variable $V_{c}(\sigma, \sigma')$ at the corresponding
entry of the matrix, that is consider the (real symmetric) 
$N$ by $N$ matrix $H(\sigma, \sigma')=V_{c}(\sigma, \sigma')$. Diagonalize it,
plot the eigenvalues, and repeat the operation a lot of times 
to get some statistics.  

We have implemented this program and run it on a computer for $N$
up to about 600 [G. Poirot, unpublished].  We cannot yet 
predict clearly scaling laws or tail estimates
from this numerical study,
but the results indicate that the eigenvalue distribution
is differently peaked 
and extends more towards large eigenvalues that in the free case.
This is not surprising, since constraints typically increase both the 
degeneracy hence the small eigenvalues of a matrix, and allow for larger 
exceptional eigenvalues. For instance the completely constrained
matrix with the same random value at every entry has $N-1$ eigenvalues
0, and a single large eigenvalue of size $N$.

We close this parenthesis on this
simple discretization of the random matrix model,
and return now to the true
problem which is complicated by spatial dependence. For this true
problem one gets 
better bounds by considering angular sectors, called {\it anisotropic},
which are wider in the tangential
than in the radial directions [FMRT1]. We introduce now
these sectors and the precise tools to prove Theorem I
below.


\subsection{Anisotropic Sector Decomposition} 
We noticed 
that the cutoff $\eta_{j}$ implies the condition:
\begin{equation}
\quad\quad\quad  \quad\quad\quad   | |p| - 1 | \le 2 M^{-j}  \ .
\end{equation}

We construct a partition of the unit sphere into anisotropic
sectors [FMRT2] $S_{\alpha}$, by projecting e.g. a standard division
of the faces of the 
cube $[-1,1]^{3}$ into $6 \times 4M^{(2j)/2}=24M^{j}$ square
plaquettes of side $M^{-j/2}$, hence area
$M^{-j}$, onto the sphere. This gives a sharp partition of the sphere
into $24M^{j}$ ``spherical plaquettes'' or sectors $\alpha$ of variable
area (between $M^{-j}$ and ${M^{-j}\over 3}$, since the maximal distance
of the cube to the origin is $\sqrt3$). We call $k_{\alpha}$ the center of
mass of $\alpha$.
From this sharp partition one can build an associated 
smooth $C^{\infty}$ scaled partition of unity $\bar\chi_{\alpha}$ on the sphere
with Urisohn's lemma. Finally we transport this partition
into a partition of the support of the cutoff $\eta_{j}$ by writing: 
\begin{equation}
\quad\quad\quad  \quad\quad\quad\eta_{j} = \sum_{\alpha} \eta_{\alpha}
\end{equation}
where $\eta_{\alpha}(p)= \eta_{j}(p) \cdot \bar\chi_{\alpha}(p/|p|)$.

The support $S_{\alpha}$ of a sector $\alpha$, or of an $\eta_{\alpha}$ function, is 
therefore included in a parallelepipedic slab, 
of center $k_{\alpha}$, which has small thickness $ l= M^{-j}$
in the direction parallel to $k_{\alpha}$
and which is longer in the two other perpendicular directions 
where it has length $L = 4 M^{-j/2} (1+{\textstyle \frac{2}{M}})=
O(1)M^{-j/2} $.

The cube and our sector partition
is invariant under parity: $p\to -p$. Therefore
for any sector $\alpha$ with center of mass $k_{\alpha}$ we can define 
the antipodal sector $\bar\alpha =-\alpha $. All other sectors have
support at distance at least say $(1/4)M^{-j/2}$ of the antipod 
center $k_{\bar \alpha}=-k_{\alpha}$, if the partition of unity $\chi_{\alpha}$
does not spread too much onto the nearest neighbors, a fact we can assume
from now on.

Then if $\beta \ne \alpha$, we define

\begin{equation}
   k_{\alpha -\beta} = k_{\alpha} - k_{\beta} = 
     2  \gamma_{\alpha-\beta} \cos x_{\alpha,\beta}  \,
\end{equation}
where $x_{\alpha,\beta}=\theta_{\alpha,\beta}/2$ is half the angle $\theta_{\alpha,\beta} $ 
(modulo $\pi$) 
between $k_{\alpha}$ and $-k_{\beta}$, so $x_{\alpha,\beta}\in [0,\pi/2[$, and 
$\gamma_{\alpha - \beta}$
is a well defined unit vector.

\subsection{Triplet of cubes operator}

It is natural to consider a smooth partition of the space 
by a lattice ${\cal D}$ 
of cubes $\Delta$ of side $M^{j}$, with 
$C_{o}^{\infty}$ ``characteristic functions'' $\chi_{\Delta}$.
There is a corresponding orthogonal decomposition
of the random variable $V= \sum_{\Delta}V^{\Delta}$,
by writing its covariance $\kappa(x,y)= \sum_{\Delta}
\kappa^{1/2}(x,z)\chi _{\Delta}(z) \kappa^{1/2}(z,y)$.

\medskip
We define $ H^{\Delta_{z}} (x,y) =\eta_{j} V^{\Delta_{z}} \eta_{j} $
and for each triplet $\vec{\Delta}= (\Delta,\Delta',\Delta'') \in {\cal D}^{3}$
or each pair $(\Delta, \Delta')\in {\cal D}^{3} $

$$
H_{\vec{\Delta}} (x,y) \equiv
H_{{\Delta''} \atop {\Delta \Delta '}}
= \chi_{\Delta}(x) H^{\Delta''} (x,y) \chi_{\Delta'} (y)$$
\begin{equation}
\quad \quad \quad \quad\quad\quad \quad  +\chi_{\Delta'}(x) H^{\Delta''} (x,y) 
\chi_{\Delta} (y) 
\end{equation}

Similarly we define 
$$
  C(\Delta, \Delta')(x,y) \equiv \chi_{\Delta}(x)  C (x,y)   \chi_{\Delta '}(y)
$$
\begin{equation}
\quad \quad \quad \quad \quad\quad \quad  + \chi_{\Delta'}(x)  C (x,y)   
\chi_{\Delta }(y)
\end{equation}
to keep the operators symmetric.


\section{Size of the operator H}
Let us repeat that because we are in dimension $d=3$, the operator
$H$ localized in a box, considered as a matrix whose entries
are the sectors of ingoing and outgoing momenta, does not, 
even approximately,
look like a classical ensemble such
as the orthogonal or Hermitian ensembles of random matrices 
which appeared in dimension 2 [P]. So 
Wigner's semi-circle law and the usual asymptotics do not apply to this 
problem. We must use a different, less accurate, method to bound the tail of 
the distribution, namely the probability that some eigenvalue becomes large.

Let $N_{j}=M^{2j}$, and 
\begin{equation}D(\vec \Delta) = (1+ M^{-j} d(\Delta, \Delta') + M^{-j} d(\Delta ', \Delta''))
\end{equation}
Our main probabilistic bound is

\medskip
\noindent{\bf Theorem II}

{\it
Let $p\ge 1$ be some integer.
There exists a constant $C_{p}>1$ such that for $A>C_{p}$
$$ {\rm \bf Prob}\{  ||H_{\vec{\Delta}}|| \ge A.D(\vec \Delta)^{-p} N^{-1/4} \} 
$$
\begin{equation}
\quad\quad\quad\quad\quad\quad\le (AD(\vec \Delta)^{2p})^{-N^{3/16}} \ .
\end{equation}
}
\medskip

The proof is quite lengthy, and is the subject of the rest of the paper.

Recall that for a compact symmetric trace class operator such as $H$
(it is easily proved trace class
because of the cutoffs), the ordinary norm is bounded by
$(tr H^{2n})^{1/2n}$ for any positive integer $n$.
Therefore

$$ {\rm \bf Prob}\{  ||H_{\vec{\Delta}}|| \ge A.D(\vec \Delta)^{-p} N^{-1/4} \} 
$$
\begin{equation}
\le ( A.D(\vec \Delta)^{-p} N^{-1/4} )^{-2n} \int d\mu (V^{\Delta''}) tr H^{2n} \ .
\end{equation}

Performing the Gaussian integration we obtain 
a sum over all Feynman graphs obtained by contracting $2n$ insertions
of $V^{\Delta''}$ on a loop of $2n$ propagators which are $\eta_{j}(\chi_{\Delta}+
\chi_{\Delta'})^{2}\eta_{j}$. $n$ is some integer that will be adjusted later.

\vskip 5cm
\centerline{\hbox{\psfig{figure=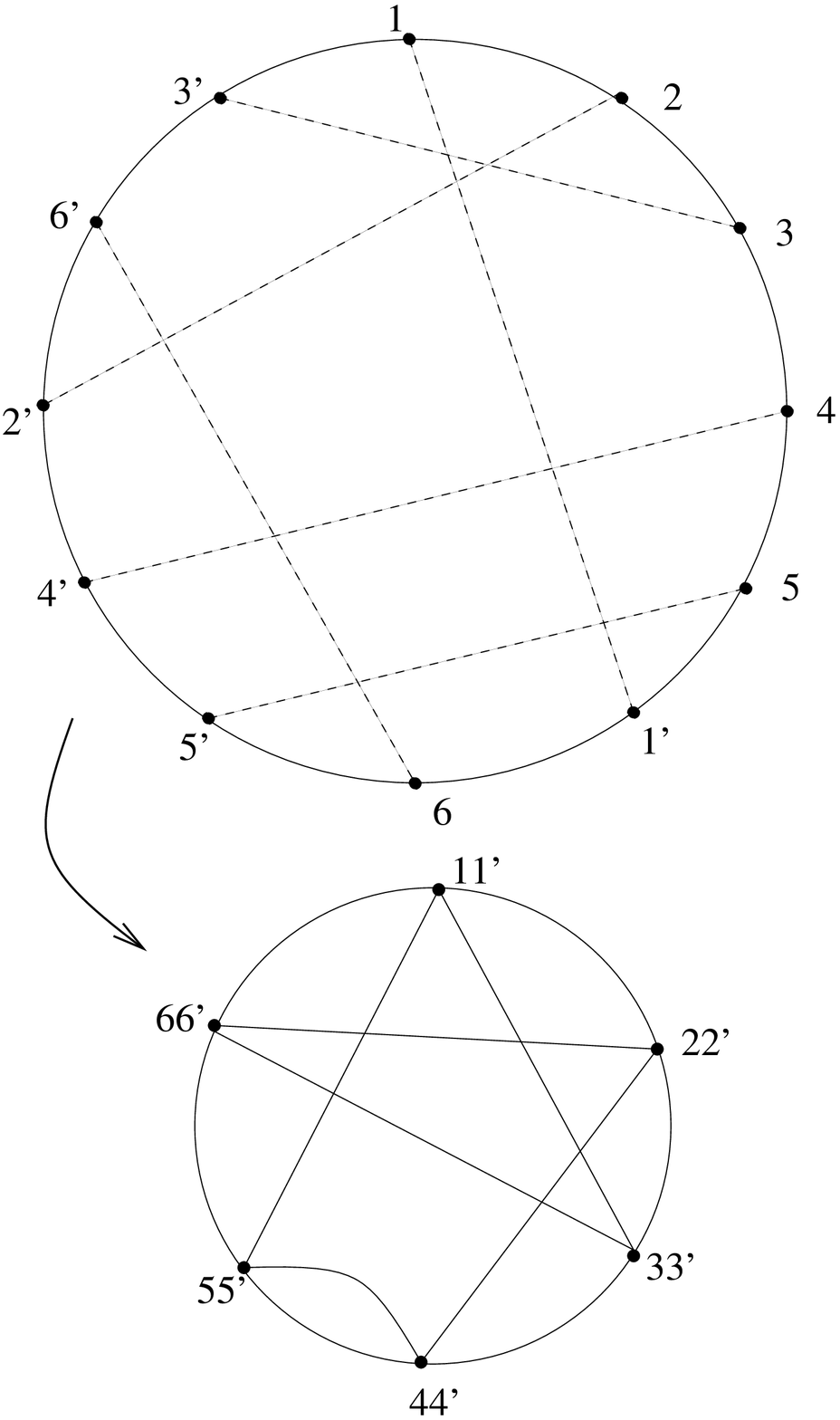,height=10cm,width=6.2cm}}}
 
\vskip 1cm
\centerline{\bf Figure 2: A graph $G$}

\centerline{\bf  and its associated
$\phi^{4}$ graph}

\centerline
{\bf obtained by contracting the dotted lines}

\vskip 1cm

 More precisely 
\begin{equation} \quad\quad\quad\quad
\int d\mu (V^{\Delta''}) tr H^{2n} =  \sum_{G\in {\cal G}_{m}} I_{G}\ ,
\end{equation}
where the amplitude for the graph $G$ is noted $I_{G}$.

To evaluate the amplitude of a given graph $I_{G}$,we split 
the spatial integrals into cubes of side size $M^{j/2}$,
hence volume $N^{3/4}$. In the language of the $\phi^{4}$ field theory,
the ``vertices'' correspond to the contractions
of the initial $V$ random field. They are propagators 
$W(x,y) = \kappa^{1/2}(x,z)\chi _{\Delta}(z) \kappa^{1/2}(z,y)$.
They are represented by the dotted lines in Figure 2. There are also
the ``ordinary lines''
which correspond to the electron propagators 
$\eta_{j} \chi_{\Delta \cup \Delta'}\eta_{j}$. They are represented by solid lines
in Figure 2.

We split these electron propagators as $\Gamma_{\Delta,\Delta'}=
\sum_{\alpha,\alpha'}\Gamma_{\Delta,\Delta'} \eta_{\alpha} \chi_{\Delta \cup \Delta'}\eta_{\alpha}$.
Since the functions $\chi_{\Delta}$ are scaled and smooth, 
we have momentum conservation, which means that $\alpha$ must be close to $\alpha'$,
and then we have spatial decay of $\Gamma$ dual to the length of the cell.
Therefore:

\medskip

\noindent{\bf Lemma 1} 

{\it
For any integer $q$, there exists a constant $C_{q}$ such that
$$ | \Gamma_{\Delta,\Delta'}^{\alpha,\alpha'}(x,y) | 
$$
$$\le {C_{q} M^{-2j}\over
\left[1+M^{-2j} \mbox{d}^{2}(x, \Delta\cup\Delta')\right]^{q} 
\left[1+M^{j} \mbox{d}^{2}(\alpha,\alpha')\right]^{q} }$$
\begin{equation}
  \frac{1}{\left[1+M^{-2j} \mbox{d}^{2}_{\alpha_{/\!/} }(x,y)\right] ^{q} 
    \left[ 1+M^{-j} \mbox{d}^{2}_{\alpha_{\perp}}(x,y)\right]^{q} } 
\end{equation}
where $d^{2}(\alpha,\alpha')$ is the distance on the sphere between sectors
$\alpha$ and $\alpha'$, and $d_{/\!/}$ and $d_{\perp}$ are the distances
in the axis parallel to the center of $\alpha$ and in the plane perpendicular
to it.}

\medskip
\noindent{\bf Proof\ } This is an easy exercise in integration by parts, 
similar to Lemma 4 in [FMRT2] (pg 689). 

\medskip
Applying also standard rules of integration
by parts (see e.g. [FMRT2]), we can extract a very small factor 
if sectors $\alpha_{1}$, $\alpha_{2}$, $\alpha_{3}$ and $\alpha_{4}$
meet at a given vertex, and do not contain points which add up to 0.
This mean that we can bound every vertex, extracting
such a momentum conservation
factor, plus some distance factor, by
$$  \frac{C_{q}} 
{\left[1+M^{j} |\bar\alpha_{1} +\bar\alpha_{2}+\bar\alpha_{3} +
\bar\alpha_{4}| \right]^{q}}
$$
\begin{equation} \quad \quad \quad \quad \quad 
{1 \over \left[1+M^{-2j} \mbox{d}^{2}(x_{v},
\Delta'') \right]^{q} }
\end{equation}
where $\bar\alpha_{i}$ is the center of the sector $\alpha_{i}$. 

Substituting these bounds we obtain our general bound for a graph of order
$n=2m$, taking into account the fact that each half vertex joins
successively $\Delta$, $\Delta'' $ and $\Delta'$ or $\Delta'$, $\Delta''$ and $\Delta$:

$$ I_{G} \le C_{q}^{3n}D(\vec \Delta)^{-qn}$$
$$
\sum_{\{\alpha_{l},l\in G\}} \int_{\Delta\cup\Delta'}dx_{1}
\int_{{\bf R}^{3}} dx_{2}... dx_{n} $$
$$\prod_{v} \frac{1} 
{\left[1+M^{j} |\alpha_{1} +\alpha_{2}+\alpha_{3} +\alpha_{4}| \right]^{q} }
$$
$$\prod_{l} \frac{N^{-1}}{
\left[1+M^{j} \mbox{d}^{2}(\alpha_{l},\alpha'_{l})\right]^{q}
\left[1+M^{-2j} \mbox{d}^{2}_{\alpha_{l, /\!/}}(x_{l},y_{l})\right]^{q}} $$
\begin{equation}\quad\quad\quad\quad\quad \frac{1}{ 
    \left[ 1+M^{-j} \mbox{d}^{2}_{\alpha_{l,\perp}}(x_{l},y_{l})\right]^{q}} 
\ . 
\end{equation}
where the sum runs over attributions of sectors to the lines, and spatial
integrations over the vertices have been performed, except at one point
per vertex.

Now by a combinatoric factor 2 per vertex we can decide  if there
exists at each vertex a sum or a difference between two incident
sectors which is small in the sense 
that $|\alpha_{v}\pm \alpha_{v}' | \le M^{-j((1/2)-a)}$.
$a$ is some small number that we will fix later to $a=1/32$.
In this way we cut the amplitude
$I_{G}$ in $2^{n}$ elements as 
$I_{G}=\sum_{t} I_{G,t}$, where the sum over $t$
runs over the choices at each vertex whether there is or not such a small sum
or difference.

Vertices such that there is no such small sum or difference are called
``twisted'' or non planar. The vertices with such a small difference
are called ``untwisted'' since their four momenta typically
will lie approximately in a common plane, or otherwise the 
vertex factor in (31) will be very small. The number of twisted
vertices is called $n_{t}$, and the number of untwisted vertices
is $n_{u}$, with $n=n_{t}+n_{u}$.

Now for each ``untwisted vertex'', we choose the particular pair
of momenta realizing the smallest sum or difference (or an arbitrary
one if several pair realize this minimum) and we split the vertex
into two half vertices and an intermediate wavy line
according to this splitting. The graph has then $n_{t}$ quartic
vertices and $n_{u}$ pairs of trilinear vertices.
Then we cut all the corresponding wavy lines. The graph splits then into
$c$ ``ordinary''
connected components which contain at least one twisted vertex, plus
possibly other twisted vertices and
chains of untwisted half vertices, 
and into $c'$ pure ``cycles'' of untwisted half vertices, with $c+c'\ge 1$.
The result of this splitting process is pictured in Figure 3, for a graph
with 15 vertices, 8 of which are untwisted and seven twisted.
It splits into $c=5$ connected components if the wavy lines (untwisted
vertices) are removed. We have $c'=3$ of these 5 components which are
cycles which do not contain twisted vertices.

We select a tree of wavy lines relating together all these
components, and will perform the spatial integration and attribution
of momentum sectors according to what we call ``painting rules''
for each connected components. The tree of wavy lines is ordered by
chosing a particular root, which correspond to a special
connected component $G_{0}$ (either ordinary or cycle).

\vskip 1cm
\centerline{\hbox{\psfig{figure=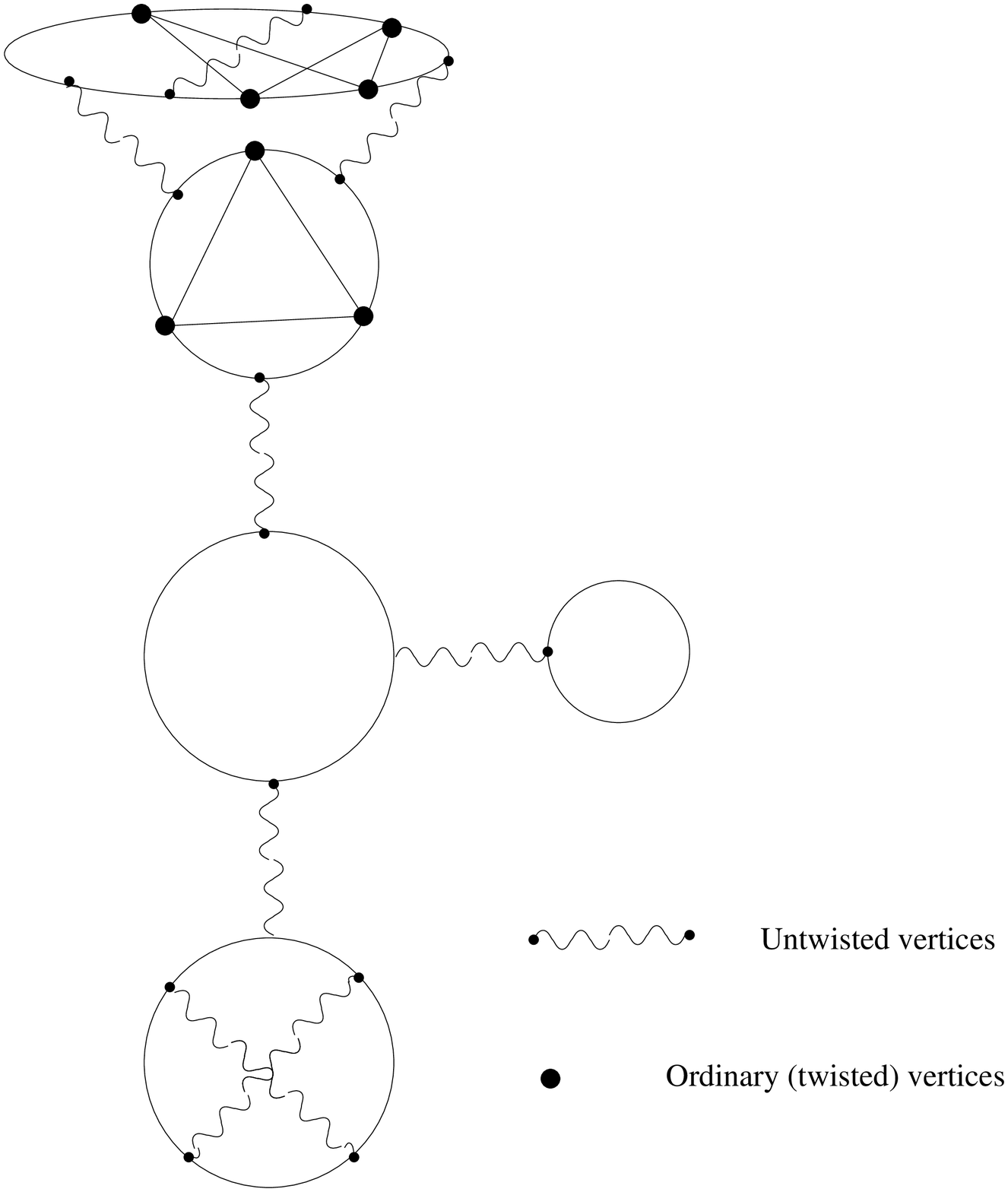,height=10cm,width=6.2cm}}}
 
\vskip .5cm
\centerline{\bf Figure 3: Splitting a graph} 

\centerline{\bf into twisted and untwisted vertices}

\vskip .5cm

The standard weight for integration of a vertex with the help
of a single propagator hooked to it is the volume
of a tube of size $M^{j/2}$ in the two
perpendicular directions to a sector, and $M^{j}$
in the parallel direction, hence 
it is $N$. However if the vertex is {\it twisted} and integrated by taking
into account the decay of two propagators hooked to it,
remark that the corresponding  weight is at most $N^{1-a/2}$,
since we gain the fact that the integration is located at the intersection of
two tubes who make an angle at least $M^{-j((1/2)-a)}$. Finally
there is one single vertex in the root component whose integration
costs the full volume of $\Delta \cup \Delta \cup \Delta''$, hence $M^{3j}=N^{3/2}$.

Now let us perform momentum attributions and vertex integrations. 
We perform these attributions in an order induced from
the tree chosen and from fixing rules associated 
to each connected components. Let $G_{k}$ be pure cycles
of  $p_{k}$ untwisted half vertices. 
Let $r_{k}$ be the number of such half-vertices
whose other half vertex is not yet fixed in the process (i.e.
is lower in the tree). We have $r_{k}\ge 1$, except possibly
for the last cycle $G_{0}$ if it is the root of the tree. 
Then the first sector attribution to a line of $G_{k}$ is made
on a half vertex of the type $r_{k}$ (except possibly if $k=0$).
It costs $N^{1/2}$
(total number of sectors). Each next attribution costs either
$N^{a}$ (number of sectors in the cone
of opening angle $M^{-j((1/2)-a)}$), 
or $O(1)$, depending whether the sectors of
the other half vertex to which this half vertex
is associated has been already attributed 
or not. So the total factor is 
$N^{(1/2)+ a (r_{k}-1)}= N^{(1/2)-a + ar_{k} }$, except possibly for
$k=0$, where it can be $N^{1/2}$ if $r_{0}=0$, so there
is an extra $N^{a}$ in that case, with respect to the formula 
$N^{(1/2)-a + ar_{k} }$.

Finally it remains to attribute the momenta in the ordinary components.
Let us contract the chains of untwisted vertices to single lines
(but with an index $p$ for the number of insertions of half untwisted
vertices, and an index $r$ for
the number of such insertions whose other
half vertex is not yet fixed). 
Fixing all the sectors of such a line is like fixing
one of them at the end, plus a factor $N^{a.r}$. 

To complete the bound, it remains therefore to compute the weight of
fixing sectors in a vacuum connected $\phi^{4}$ graph solely made
of {\it twisted} vertices. Let $\bar G$ be such a graph, and $s$
its number of vertices.

A fixing rule $P$ on $\bar G$ is an ordering of the vertices of $\bar G$
as $v_{1},...,v_{s}$ such that for every $k>1$ there exists a line hooked 
at one end to $v_{k}$ and at the other end to one vertex
among $\{v_{1},...,v_{k-1}\}$. There always obviously exist such 
fixing rules. For instance to any tree $T$ of $G$ and any way of
``turning around the tree'' (i.e. choice of a root and relative
ordering of the non trivial forks going up in the tree) is associated a
fixing rule corresponding to the ordering in which vertices of
$G$ are met by ``climbing around the tree'' in that way. 

The $k$-th vertex $v_{k}$ for the fixing rule $P$ 
is called of type 4, 3, 2, 1 or 0 with respect to $P$
if the number of half lines hooked at one end to $v_{k}$ and at the other
end to a vertex not among $\{v_{1},...,v_{k-1}\}$ is 4,3,2,1, or 0.
The number of vertices of type $r$ is called $n_{r}$. 
Obviously $n_{4}+n_{3}+n_{2}+n_{1}+n_{0}=s$, and
$n_{4}=1$ since the first vertex of the fixing rule is the only one
of type 4. 

To perform the sum over the values of sectors in the graph
we fix first the indices of vertex $v_{1}$, which is the only one of type
4. This costs $N^{1/2}N^{1/2}N^{1/4}=N^{5/4}$.
Then vertices of type 3 cost $N^{1/2}N^{1/4}= N^{3/4}$.
Vertices of type 2 cost $N^{1/4}$, because they are twisted. 
Vertices of type 1 or zero 
cost nothing. Finally we should take 
into account the fact that because
of the twisting, vertices of type 2,1 or 0, when integrated, 
are integrated with the decay of two propagators whose sectors are known and
have a minimum angle. This allows a gain of $N^{-a}$, as explained above,
for the spatial integration of these vertices.

Collecting all factors, the key point is to remark that the
factors $N^{ar_{k}}$ and $N^{ar}$
combine into $N^{an_{u}}$, since
one of the two half vertices of any untwisted vertex
is always fixed before the other. Adding the extra $N^{a}$
eventual factor for $G_{0}$, 
we arrive at the following estimate, where we recall that
$n_{u}$ is the total number of untwisted vertices:

$$ I_{G,t} \le C_{q}^{3n}D(\vec \Delta)^{-qn} N^{-2n} N^{n+1/2} N^{a(n_{u}+1)}
 $$
\begin{equation}
N^{(1/2-a)c' + (5/4)c+(3/4)n_{3} + ((1/4)-a)n_{2}
-a(n_{1}+n_{0})} 
\end{equation}
where we have $n=n_{u}+c+n_{3} +n_{2}+n_{1}+n_{0}$, and $2n= 2n_{u} +
 4c +3n_{3} +2n_{2}+n_{1}$, counting respectively the vertices and the lines.
Taking into account the second relation we find
$$ I_{G,t} \le C_{q}^{3n}D(\vec \Delta)^{-qn} N^{1/2-n/2} N^{c'/2-n_{u}/2}
$$
\begin{equation}
N^{c/4 -n_{2}/4 - n_{1}/4 }N^{a(n_{u} 
+1-(n_{0}+n_{1}+n_{2}+c'))}
\end{equation}

Taking $q=4p$, the factor $N^{-n/2}D(\vec \Delta)^{-qn/2}$ 
kills the $N^{n/2}D(\vec \Delta)^{2pn} $ in formula (27). It remains:

$$ N^{n/2}D(\vec \Delta)^{2pn}|I_{G,t} | = J_{g,t}
$$
$$\le K_{q}^{n}D(\vec \Delta)^{-2pn} N^{a+1/2}N^{c'/2}$$
\begin{equation} 
N^{-n_{u}/2 + c/4 -n_{2}/4 - n_{1}/4 }N^{an_{u} -a(n_{0}+n_{1}+n_{2}+c')}
\end{equation}
From the fact that there is a tree of untwisted ``wavy lines'' between
the connected components of type $c$ and $c'$, we know that 
$c+c' \le 1 +n_{u}$. Therefore
$$ J_{g,t}
\le K_{q}^{n}D(\vec \Delta)^{-2pn} N^{a+3/4} $$
\begin{equation}
N^{-(1/4-a)(n_{u}-c')-n_{2}/4 - n_{1}/4}N^{-a(n_{0}+n_{1}+n_{2})}
\end{equation}
Choosing a=1/16, we have:
$$ J_{g,t}
\le K_{q}^{n}D(\vec \Delta)^{-2pn} N^{a+3/4} $$
\begin{equation}\quad\quad\quad
N^{-(3/16)(n_{u}-c') -(1/16)(n_{0}+n_{1}+n_{2})}
\end{equation}

We have  $n_{t}=c+n_{3} +n_{2}+n_{1}+n_{0}$, and $2n_{t}= 
4c +3n_{3} +2n_{2}+n_{1}$, therefore $2c + n_{3} = n_{1} +2n_{0}$.
Therefore $c+n_{3} \le n_{1} +2n_{0}$, so $n_{t} \le n_{2}+2n_{1}+3n_{0}
\le 3(n_{2}+n_{1}+n_{0})$. Hence
$$ J_{g,t}
\le K_{q}^{n}D(\vec \Delta)^{-2pn} N^{a+3/4} $$
$$
N^{-(3/16)[(n_{u}-c')-(3/16)n_{t}]}$$
\begin{equation}\quad\le K_{q}^{n}D(\vec \Delta)^{-2pn} N^{a+3/4} 
N^{-(3/16)(n-c')}
\end{equation}

Now we consider the following reduction process. We contract a certain number
$x\le c'$ of pure cycles in the following inductive way:  
first we contract every tadpole (pure cycle with $p=1$),
i.e. suppress it together with one
of the two external lines hooked to the tadpole. Then we repeat this operation 
inductively (since the contraction at first stage can create new tadpoles),
until complete exhaustion. In the case of the graph of Figure 3,
there is a single tadpole so the operation stops after one step,
but in general it can be more complicated (see Figure 4).

\vskip 1cm
\centerline{\hbox{\psfig{figure=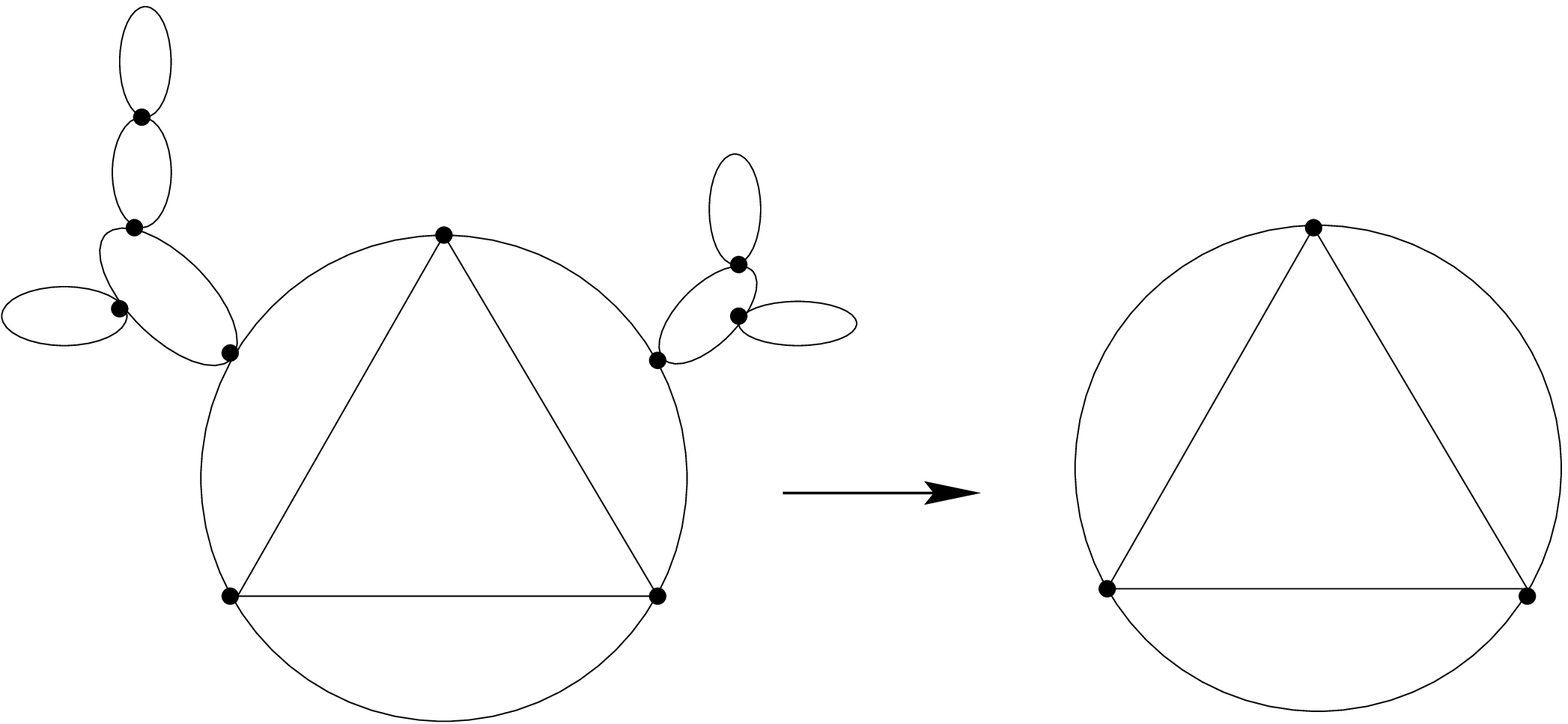,height=3.4cm,width=7cm}}}
 
\vskip .5cm
\centerline{\bf Figure 4: The contraction of tadpoles}

\vskip 1cm
\centerline{\hbox{\psfig{figure=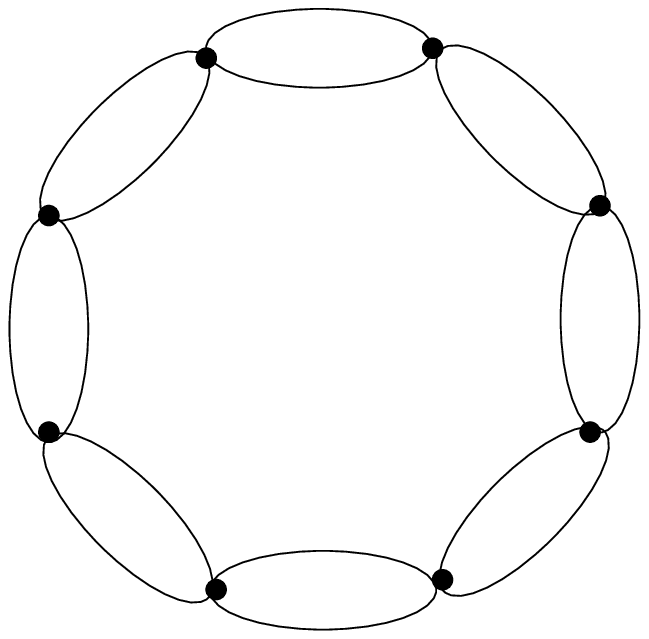,height=5cm,width=5cm}}}

\vskip .5cm 
\centerline{\bf Figure 5: a pure loop of 8 bubbles}
\vskip 1cm
 
Let $x_{1}$ be the number of cycles contracted
in this way. This contraction process continues until either we arrive
at the empty graph because at some
stage we obtained a graph with $c'=2$ made of two tadpoles, 
which we then delete 
or until we arrive at
a graph with $c'-x_{1}$ cycles without any tadpole. In this last case
either the graph is a pure loop of $x_{2}$ bubbles that we destroy (see Figure
5),
or it is not. In this second case,
the graph may still contain pure bubbles, i.e.
loops without twisted vertices of length 2.
For instance in the case of the graph of Figure 3, after contraction
of the tadpole, there is one such bubble which appears.

We perform then a second operation of contraction,
which contracts every maximal chain of such bubbles
(pure cycles of length 2 without twisted vertices) to a single 
new type of wavy line.
This second operation has not to be repeated inductively in contrast
with the first, since it cannot regenerate any bubble
(see Figure 6). (Wavy lines have been omitted in Figure 4-5-6
for simpler pictures). 

\vskip 1cm
\centerline{\hbox{\psfig{figure=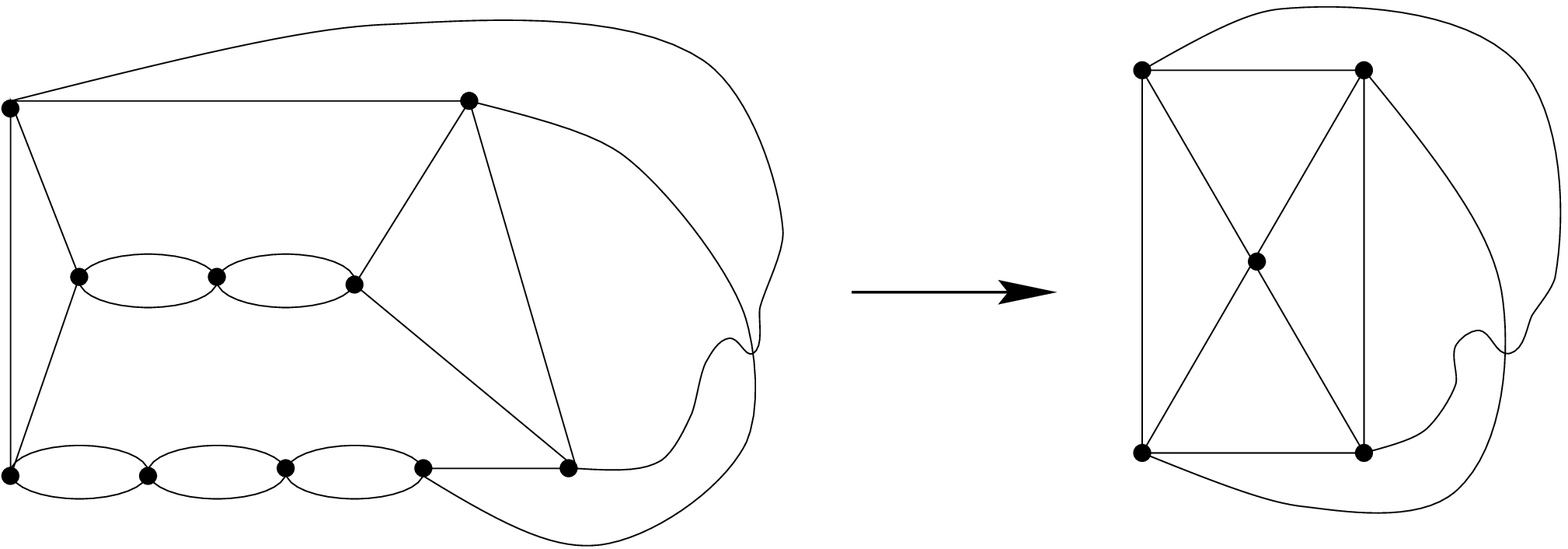,height=2.8cm,width=6.5cm}}}

\vskip .5cm 
\centerline{\bf Figure 6: the reduction of bubbles}
\vskip1cm

Let $x_{2}$
be the total number of bubbles contracted in this way. 
We arrive at a graph with $c'-x_{1}-x_{2}$ cycles, each of them having at 
least three lines. 

Let $x=x_{1}+x_{2}$.

\medskip
\noindent{\bf Lemma 2}
\noindent{\it We have $x=x_{1}+x_{2} \le f_{2}(G) + 2$, where $f_{2}(G)$
is the cardinal of the largest 
forests of connected 4-point subgraphs of $G$, defined in [CR].}

Proof: Obviously each contracted cycle can be associated to a new
4 point subgraph in a forest, except if $x=c'$, in which case
when one destroys the two tadpoles there is no new 4-point
subgraph associated, or when one destroys the loop of $k$ bubbles,
there is only $k-1$ corresponding 4-point subgraphs in a forest.
This two cases are exclusive, which explains the 2 in the lemma.

We have obviously $x\le c'$ hence $n-c' \ge n -x \ge n-f_{2}(G)-2$.
We use now:

\medskip
\noindent{\bf Lemma 3 (particular case of [CR], theorem II)}
\noindent{\it
The number of Wick contractions of half-vertices 
giving rise to a vacuum connected
graph with $n$ vertices and $f_{2}(G)+2=f$, 
is bounded by $K^{n} n!/f! \le K^{n}n^{n-f}$, for some constant $K$.}

\medskip
\noindent{\bf Proof}\  See ([CR], Theorem II).

Combining this, with the previous bound we obtain (since $f \le n+2$ ([CR])):
$$ \sum_{G} J_{G} \le $$
$$\sum_{f\le n + 2} (KK_{q})^{n} n^{n-f} 
D(\vec \Delta)^{-2pn} N^{a+3/4} 
N^{-(3/16)(n-f)}$$
\begin{equation} \le
C^{n}  D(\vec \Delta)^{-2pn} N^{a+3/4} (n/N^{3/16})^{n-f}
\end{equation}
where $C=2 K.K_{q}$ is a constant. 
We choose now $n=N^{3/16}$
and obtain:

$$ {\rm \bf Prob}\{  ||H_{\vec{\Delta}}|| \ge A.D(\vec \Delta)^{-p} N^{-1/4} \} 
$$
\begin{equation}\quad\quad
\le A^{-2N^{3/16}} D(\vec \Delta)^{-2pN^{3/16} }  N^{a+3/4} 
C^{N^{3/16}} 
\end{equation}

If $A^{1/2}$ is larger than $C$, we obtain easily the bound (32) of the theorem
(since $A^{-N^{3/16}/2} N^{a+3/4} \le 1 $, $A$ being large and $N\ge 1$,
and we can also choose $A$ such that $CA^{-1/2} \le 1$).
We have therefore in that case:

$$ {\rm \bf Prob}\{  ||H_{\vec{\Delta}}|| \ge A.D(\vec \Delta)^{-p} N^{-1/4} \} 
$$
\begin{equation}\quad\quad\quad\quad\quad\quad
\le A^{-N^{3/16}} D(\vec \Delta)^{-2pN^{3/16} } 
\end{equation}
which completes the proof of Theorem II. \endproof

In order to obtain a theorem no longer on $H$ but on the full
operator $\lambda C H$ that occurs in our resolvent expansions,
we need to add to the estimate of Theorem I the power counting coming from
the coupling constant and the norm of the propagator.
This adds a factor $\epsilon N^{1/4}$ to the norm of $H$. The factor
$\epsilon$ comes from the distance to the last slice, since
recall that in section 3 we assumed that $\lambda^{2} < \epsilon M^{-j}$.
Finally we obtain, taking $\epsilon=A^{-2}$ small, the following estimate:  

\medskip
\noindent{\bf Theorem III}

{\it
If the coupling is kept small enough (i.e. $\epsilon \le A^{-2}$,
where $A$ is defined by Theorem II) we have:

$$ {\rm \bf Prob}\{  ||\lambda C H_{\vec{\Delta}}|| 
\ge A^{-1} D(\vec \Delta)^{-p} \}$$
\begin{equation}
\quad\quad\quad\quad\quad\quad
\le A^{-N^{3/16}} D(\vec \Delta)^{-2pN^{3/16} }  
\end{equation}
}
\medskip

To complete the proof of Theorem I, one combines the estimate of Theorem III
with a large versus small
field cluster expansion. This expansion is exactly similar to the one
of [P], to which we refer the reader. The decay factor 
$D(\vec \Delta)^{-p}$ is roughly speaking needed to sum the various
triplets $\vec \Delta$ with respect to one of their elements,
and the factor $A^{-N^{3/16}}$ is sufficient to
compensate the bad factors due to the imaginary translation
(see (13)) which come from the estimation
of the resolvent in the large field regions.

\medskip
\noindent{\bf Acknowledgements}
\medskip

We thank J. Feldman, H. Kn\"orrer and E. Trubowitz for many 
fruitful discussions and for their interest in this program. 
We also thank A. Connes for pointing reference [V] to us,
and Wei-Min-Wang for an interesting discussion.

\medskip
\noindent{\bf References}
\medskip

[A] M. Aizenman, Localization at Weak Disorder: Some Elementary Bounds,
in {\it The State of Matter}, edited by M. Aizenman and H. Araki, World
Scientific 1994.

[AM] M. Aizenman and S. Molchanov, Localization at large
disorder and at extreme energies: an elementary derivation. Commun. Math.
Phys. 157, 245 (1993).

[CR] C. de Calan and V. Rivasseau,
Local existence of the Borel transform in Euclidean $\phi^{4}_{4}$, 
Commun. Math. Phys. 82, 69 (1981).

[DLS] F. Delyon, Y. L\'evy and B. Souillard, Commun. Math.
Phys 100, 463-470 (1985).

[FMSS] J. Fr\"ohlich, F. Martinelli, E. Scoppola and T. Spencer,
Commun. Math. Phys. 101, 21-46 (1985).

[FS] J. Fr\"ohlich and T. Spencer, Absence of diffusion in the Anderson
tight binding model for large disorder or low energy, Commun. Math.
Phys. 88, 151-184 (1983).

[FT] J. Feldman and E. Trubowitz, The Flow of an Electron-Phonon System
to the Superconducting State, Helvetica Physica Acta 64 214-357 (1991).

[FMRT1] J. Feldman, J. Magnen, V. Rivasseau
and E. Trubowitz, A Rigorous Analysis of the Superconducting Phase
of an Electron-Phonon System, Proceedings
of Les Houches Summer school 1994 (F. David, P. Ginsparg eds).

[FMRT2]  J. Feldman, J. Magnen, V. Rivasseau
and E. Trubowitz, 
An Infinite Volume Expansion for Many Fermion Green's functions,
Helv. Phys. Acta, Vol. 65, 679 (1992).

[FMRT3]  J. Feldman, J. Magnen, V. Rivasseau
and E. Trubowitz, An Intrinsic 1/N Expansion for Many Fermion Systems,
 Europhys. Letters 24, 437 (1993).

[FMRT4]  J. Feldman, J. Magnen, V. Rivasseau
and E. Trubowitz,
Two Dimensional Many Fermion Systems as Vector Models,
 Europhys. Letters 24, 521 (1993).

[K] A. Klein, Absolutely continuous spectrum in the Anderson model 
on the Bethe lattice,
 Mathematical Research Letters {\bf 1}, 399-407 (1994).

[MPR]  J. Magnen, G. Poirot and V. Rivasseau, in preparation.

[P] G. Poirot, ``A Single Slice 2d Anderson Model at Weak Disorder'', preprint 
cond-mat/9605140.

[R] V. Rivasseau, Cluster expansions with large/small field conditions,
Vancouver summer school 1993, CRM Proceedings volume 7, 1994.

[S] T. Spencer, The Schr\"odinger equation with a random potential, 
a mathematical review, in {\it Critical Phenomena, Random Systems, Gauge 
Theories}, Les Houches XLIII, ed. by K. Osterwalder and R. Stora.

[V] D. Voiculescu, Limit Laws for Random Matrices and Free Products, 
Invent. Math. 104 201-220 (1991). 

\end{document}